\documentclass[12pt,showpacs,preprintnumbers,superscriptaddress,amsmath,amssymb,nofootinbib]{revtex4}
\usepackage{graphicx}
\usepackage{dcolumn}
\usepackage{bm}
\usepackage{amssymb}
\usepackage{amsmath}
\usepackage{epsfig}    
\usepackage{color}
\usepackage{slashed}
\usepackage{hhline}

\def\be{\begin{equation}}
\def\ee{\end{equation}}
\newcommand{\bea}{\begin{eqnarray}}
\newcommand{\eea}{\end{eqnarray}}
\newcommand{\nn}{\nonumber}

\numberwithin{equation}{section}

\begin{document}

\begin{flushright}
KIAS-P13067\\
IPPP/13/100\\
DCPT/13/200
\end{flushright}


\title{Two Loop Neutrino Model and Dark Matter Particles with Global $B-L$ Symmetry}

\author{Seungwon Baek}
\email{swbaek@kias.re.kr}
\affiliation{School of Physics, KIAS, Seoul 130-722, Korea}
\author{Hiroshi Okada}
\email{hokada@kias.re.kr}
\affiliation{School of Physics, KIAS, Seoul 130-722, Korea}
\author{Takashi Toma}
\email{takashi.toma@durham.ac.uk}
\affiliation{Institute for Particle Physics Phenomenology University of
Durham, Durham DH1 3LE, UK}


\begin{abstract}
We study a two loop induced seesaw model with
global $U(1)_{B-L}$ symmetry, in which we consider two component dark matter
 particles. 
The dark matter properties are investigated together with some phenomenological
 constraints such as electroweak precision test, neutrino masses and
 mixing and lepton flavor violation. 
In particular, the mixing angle between  the Standard Model like Higgs
 and an extra Higgs is extremely restricted by the
 direct detection experiment of dark matter. 
We also discuss the contribution of Goldstone boson to the effective
 number of neutrino species $\Delta N_{\rm eff}\approx0.39$ which has
 been reported by several experiments. 
\end{abstract}


\maketitle\newpage

\section{Introduction}
Even after the discovery of the standard model (SM) Higgs boson, there
still exist  some unsolved issues: the origin of neutrino masses and
mixings, 
the nature of dark matter (DM), whether Higgs boson is the only
elementary scalar particle 
or not, and so on. As for the neutrinos, the
tiny mass scale is apparently different from the other sectors, {\it
i.e.} the charged leptons 
and quarks. 
Hence many physicists believe there exist some mechanisms for neutrino
mass generation 
which is different from the other fermion mass generation. 
One of elegant solutions is to generate the neutrino masses with
radiative correction, 
which provides more natural explanation of its smallness. Moreover
neutrinos often 
interact with some new mediating particles that can be frequently
identified to be 
DM. Such kind of models have  been proposed by many authors in
ref.~\cite{Ma:2006km, Aoki:2013gzs, Dasgupta:2013cwa, Krauss:2002px, Aoki:2008av, Schmidt:2012yg,
Bouchand:2012dx, Ma:2012ez, Aoki:2011he, Ahn:2012cg, Farzan:2012sa,
Bonnet:2012kz, Kumericki:2012bf, Kumericki:2012bh, Ma:2012if, Gil:2012ya,
Okada:2012np, Hehn:2012kz, Dev:2012sg, Kajiyama:2012xg, Okada:2012sp,
Aoki:2010ib, Kanemura:2011vm, Lindner:2011it, Kanemura:2011mw,
Kanemura:2012rj, Gu:2007ug, Gu:2008zf, Gustafsson, Kajiyama:2013lja, Kajiyama:2013sza,
Hernandez:2013dta, Hernandez:2013hea, McDonald:2013hsa, Okada:2013iba}. 

As for DM, its properties are being explored by various experiments
such as direct detection and indirect detection experiments as well as
Large Hadron Collider (LHC). 
For example, the current direct detection experiment LUX~\cite{xenonlux}
tells us that the upper bound for the spin 
independent cross section is highly constrained to 
be ${\cal O}(10^{-46})\  {\rm cm^2}$ at 
around $50$ GeV of DM mass.
For indirect detection, the recent analysis of Fermi-LAT gamma-ray
data has shown that there
may be gamma-ray line peak near 130 GeV, which could be interpreted as 
annihilation or decay of DM~\cite{Bringmann:2012,
Weniger:2012tx, Fermi-LAT:2013uma, Baek:2013ywa}.
AMS-02 experiment also has shown the anomaly in the positron fraction up to 
energy about 350 GeV, and its result is in good agreement with the previous PAMELA
experiment~\cite{ams-02, Adriani:2008zr}. They also suggest that leptophilic
DM~\cite{Fox:2008kb, Baek:2008nz, Kyae:2009jt, Ibarra:2009bm, Chun:2009zx, Das:2013jca}
 is preferable since PAMELA has reported the anti-proton-to-proton ratio
 which is consistent with the predicted background~\cite{Adriani:2008zq}. 

In this paper, we construct a two-loop radiative seesaw model with 
global $B-L$ symmetry at the TeV scale based on the
paper~\cite{Kajiyama:2013rla}\footnote{See for example the recent works
on local $B-L$ symmetries in non-supersymmetric
theory~\cite{Abdallah:2011ew, Abdallah:2012nm, Bandyopadhyay:2012px,
Datta:2013mta, Chakrabortty:2013zja}.}. 
We also analyze the multi-component DM properties, 
and  we discuss their detectability in addition to the observed relic
density~\cite{relic-1, relic-2}. 
At the end, we discuss the discrepancy of the effective number
of neutrino species $\Delta N_{\rm eff}\approx0.39$ between theory and
experiments which is recently suggested
by ref.~\cite{weinberg}. 

This paper is organized as follows. In Section \ref{sec2}, we show our model and
discuss the Higgs sector including the Higgs potential,  
its stability condition, S-T parameters and neutrino mass in the lepton sector.
We analyze DM phenomenology in Section \ref{sec3} and the differences
from our previous paper are summarized in Section \ref{sec4}. Then
finally we conclude in Section \ref{sec5}.

\section{The Two-loop Radiative Seesaw Model}
\label{sec2}
\subsection{Model Setup}

\begin{table}[thbp]
\centering {\fontsize{10}{12}
\begin{tabular}{|c||c|c|c|c|c|c|c|c||}
\hline ~~Particle~~ & ~~$Q$~~ & ~~$u^c$ & $ d^c $~~ & ~~$L$~~ & ~~$e^c$~~ & ~~ $N^c$~~
  & $S$   \\\hhline{|=#=|=|=|=|=|=|=|}
$(SU(2)_L,U(1)_Y)$ & $(\bm{2},1/6)$ & $(\bm{1},-2/3)$ & $(\bm{1},1/3)$ &
		 $(\bm{2},-1/2)$ & $(\bm{1},1)$  & $(\bm{1},0)$  &
			     $(\bm{1},0)$ \\\hline 
$U(1)_{B-L}$ & $1/3$ & $-1/3$ & $-1/3$ & $-1$ & $1$ & $1$  & $-1/2$ \\\hline
$\mathbb{Z}_2$ & $+$ & $+$ & $+$ & $+$ & $+$  & $-$  & $-$\\
\hline
\end{tabular}%
} \caption{The particle contents and the charges for fermions. 
}
\label{tab:b-l1}
\end{table}

\begin{table}[thbp]
\centering {\fontsize{10}{12}
\begin{tabular}{|c||c|c|c|c|}
\hline ~~Particle~~ & ~~$\Phi~~ $& ~~$\eta~~ $ & $\chi$ & ~~$\Sigma$~~ \\\hhline{|=#=|=|=|=|}
$(SU(2)_L,U(1)_Y)$ &  $(\bm{2},1/2)$ & $(\bm{2},1/2)$ & $(\bm{1},0)$ & $(\bm{1},0)$ 
\\\hline
$Y_{B-L}$ & $0$ & $0$ & $-1/2$ &  $1$ \\\hline
$\mathbb{Z}_2$ & $+$ & $-$ & $+$  & $+$\\\hline
\end{tabular}%
} \caption{The particle contents and the charges for bosons. 
}
\label{tab:b-l2}
\end{table}

We revisit a two-loop radiative seesaw model ~\cite{Kajiyama:2013rla}
but with global $B-L$ symmetry\footnote{See another example in
ref.~\cite{Lindner:2011it}.}. 
We add three right-handed neutrinos $N^c$, three SM gauge singlet fermion $S$,  
a $SU(2)_L$ doublet scalar $\eta$ and $B-L$ charged scalars $\chi$ and
$\Sigma$ to the SM particles\footnote{Note that several right-handed neutrinos $N_i^c$ and SM
gauge singlet fermions $S_i$ are needed to induce the neutrino masses and
mixing, but we do not have gauge anomaly problem. Adding
multi-right-handed neutrinos are one of the minimal 
requirements to obtain the observed neutrino masses and mixing. Another
way is to introduce two $SU(2)$ doublet inert bosons, see {\it e.g.} 
ref.~\cite{Hehn:2012kz}.}.
We do not need to add any $\overline{S}$ to avoid the gauge
anomaly problem as in ref.~\cite{Kajiyama:2013rla} because of the global
$B-L$ symmetry. Thus the particle contents are more economical 
as shown in Tab.~\ref{tab:b-l1} and  
\ref{tab:b-l2}. 
The $\mathbb{Z}_2$ parity is also imposed as Tab.~\ref{tab:b-l1}
and \ref{tab:b-l2} so as to forbid the type-I seesaw mechanism. As a consequence,
the parity-odd particles $N^c$, $S$, and $\eta$ can be DM candidates. 
The right handed neutrino $N^c$ is naturally 
lighter than $S$ and $\eta$ because its mass is generated at one-loop level.
The lightest one is stabilized by $\mathbb{Z}_2$ parity. 
The $\mathbb{Z}_6$ symmetry remains after the $B-L$ spontaneous
breaking, and the $\mathbb{Z}_6$ charge of each particle is mathematically defined as
$6(B-L)~\mathrm{mod}~6$. The $\mathbb{Z}_6$ charge of $S$ and
$\chi$ is $3$, and they can be called as odd particle under
$\mathbb{Z}_6$ symmetry since their transformation is same with
$\mathbb{Z}_2$ symmetry. 
Thus the stability of $\chi$ and $S$ is assured by a remnant $\mathbb{Z}_6$
parity after the $B-L$ spontaneous breaking~\cite{Kajiyama:2013rla}.
Although the remnant symmetry would be regarded as $\mathbb{Z}_2$
symmetry in a narrow meaning of the renormalizable model, the larger
$\mathbb{Z}_6$ symmetry should be taken into account if higher
dimensional operators such as $QQQL$ are considered. 

The gauge invariant and renormalizable Lagrangian for Yukawa sector and
Higgs potential are given by
\begin{eqnarray}
\mathcal{L}_{\mathrm{Y}}
&=&
(y_{\ell})_{\alpha\beta} \Phi^\dag L_\alpha e^c_\beta 
+ (y_{\nu})_{\alpha
i}L_\alpha \eta N^c_i  + (y_N)_{ij} N^c_i \chi S_j +
(y_S)_{ij}\Sigma S_iS_j  
+\mathrm{h.c.},
\label{HP0}\\
\mathcal{V}
&=& 
 m_1^{2} \Phi^\dagger \Phi + m_2^{2} \eta^\dagger \eta  + m_3^{2}
 \Sigma^\dagger \Sigma  + m_4^{2} \chi^\dagger\chi 
 + m_5 [\chi^2 \Sigma + {\rm h.c.}]  \nonumber \\ 
&&
+\lambda_1 (\Phi^\dagger \Phi)^{2} + \lambda_2 
(\eta^\dagger \eta)^{2} + \lambda_3 (\Phi^\dagger \Phi)(\eta^\dagger \eta)
+ \lambda_4 (\Phi^\dagger \eta)(\eta^\dagger \Phi)
+\lambda_5 [(\Phi^\dagger \eta)^{2} + \mathrm{h.c.}]\nn\\
&&+\lambda_6 (\Sigma^\dagger \Sigma)^{2} + \lambda_7  (\Sigma^\dagger \Sigma)(\Phi^\dagger \Phi)
+ \lambda_8  (\Sigma^\dagger \Sigma) (\eta^\dagger \eta)+\lambda_9 (\chi^\dagger \chi)^{2}\nn\\
&& + \lambda_{10} (\chi^\dagger \chi)(\Phi^\dagger \Phi)
+ \lambda_{11} (\chi^\dagger \chi) (\eta^\dagger \eta) + \lambda_{12} (\chi^\dag\chi)(\Sigma^\dag\Sigma) ,
\label{HP}
\end{eqnarray}
where the indices $\alpha$, $\beta$, $i,j=1-3$.
We assume all the parameters are  real\footnote{If the parameters are allowed to
be complex in general case, we may have darkogenesis similar to~\cite{Shelton:2010ta}.}. 
The quartic couplings $\lambda_1$, $\lambda_2$, $\lambda_6$ and
$\lambda_9$ have to be positive to stabilize the Higgs potential.  
While the scalars $\eta$ and $\chi$ are assumed not to have a vacuum
expectation value (VEV), the $B-L$ charged scalar $\Sigma$ has the VEV $\langle \Sigma
\rangle=v'/\sqrt{2}$ and is the source of the spontaneous 
global $B-L$ breaking. 
The VEV of $\Sigma$ gives the masses to the singlet $S$. 
The active neutrino masses are obtained through two-loop
level~\cite{Kajiyama:2013rla}. 
In general, we can choose a diagonal base of $y_S$ and
mass matrix of the right-handed neutrinos after the symmetry breaking.

\subsection{Higgs Potential}
After the global $B-L$ and electroweak symmetry breaking,
the scalar particles in the model mix each other and we need to rewrite
them by mass eigenstates. 
Since the particle content for scalar in the model is same with that of
ref.~\cite{Kajiyama:2013rla}, the discussion of the potential is exactly same with the
reference except the existence of the Goldstone boson $G$. 
The $\Phi^0$ and $\Sigma$ are given by 
\begin{equation}
\Phi^0=\frac{v+\phi^0(x)}{\sqrt{2}},\qquad
\Sigma=\frac{v'+\sigma(x)}{\sqrt{2}}e^{i G(x)/v'}.
\end{equation}
and they are rewritten by the mass eigenstates $h$ and $H$ as 
\begin{equation}
\left(
\begin{array}{c}
\phi^0\\
\sigma
\end{array}
\right)=\left(
\begin{array}{cc}
\cos\alpha & \sin\alpha\\
-\sin\alpha & \cos\alpha
\end{array}
\right)\left(
\begin{array}{c}
h\\ H
\end{array}
\right). 
\end{equation}
The mass eigenstates of the other scalar particles are $\eta^+$,
$\eta_R$, $\eta_I$, $\chi_R$ and $\chi_I$ where $\eta_R$ and $\eta_I$
are the real and imaginary parts of $\eta^0$, and $\chi_R$ and $\chi_I$
are the real and imaginary parts of $\chi$. 
Their masses are expressed by $m_\eta$, $m_{\eta_R}$, $m_{\eta_I}$,
$m_{\chi_R}$ and $m_{\chi_I}$ respectively. 
More detail is referred the ref.~\cite{Kajiyama:2013rla}. 
The requirements to obtain the proper vacuum
$\langle\phi\rangle\neq0$, $\langle\eta\rangle=\langle\chi\rangle=0$,
$\langle\Sigma\rangle\neq0$ also have been discussed in
the reference.

\subsection{Constraints}
There are some constraints we have to take into account. 
First, the radiative correction to gauge boson masses in the SM is
constrained by the electroweak precision tests~\cite{Barbieri:2006dq,
Peskin:1991sw, Baak:2012kk}. The constraint is 
expressed by S and T parameters, and can be rewritten 
in terms of a mass relation
between neutral and charged component of $\eta$ in the model. 
It is approximately given by 
\begin{equation}
\sqrt{\left(m_\eta-m_{\eta_R}\right)
\left(m_\eta-m_{\eta_I}\right)}\lesssim133~\mathrm{GeV}.
\label{st-cond}
\end{equation}
as have discussed in ref.~\cite{Kajiyama:2013rla}. 

Second, the constraint from neutrino masses and mixing is taken into
account. The neutrino mass matrix is derived at two loop level and written as 
\begin{equation}
\left(m_\nu\right)_{\alpha\beta}=
\sum_{i=1}^3\left(y_{\nu}^Ty_N^*\right)_{\alpha i}
\Lambda_i
\left(y_{\nu}^Ty_N^*\right)_{\beta i},
\label{neu-mass}
\end{equation}
where the loop function $\Lambda_i$ is defined as 
\begin{equation}
\Lambda_i=
\frac{m_{Si}}{4(4\pi)^4}
\int_0^1dx\int_0^{1-x}dy\frac{1}{x(1-x)}
\Biggl[
I\left(m_{Si}^2,m_{RR}^2,m_{RI}^2\right)
-I\left(m_{Si}^2,m_{IR}^2,m_{II}^2\right)
\Biggr],
\end{equation}
with
\begin{eqnarray}
I(m_1^2,m_2^2,m_3^2)&=&
\frac{m_1^2 m_2^2\log\left(\displaystyle\frac{m_2^2}{m_1^2}\right)
+m_2^2 m_3^2\log\left(\displaystyle\frac{m_3^2}{m_2^2}\right)+
m_3^2 m_1^2\log\left(\displaystyle\frac{m_1^2}{m_3^2}\right)}
{(m_1^2-m_2^2)(m_1^2-m_3^2)}
,\\
m_{ab}^2&=&\frac{ym_{\eta_{a}}^2+xm_{\chi_{b}}^2}{x(1-x)}\quad
(a,b=R\:\mathrm{or}\:I).
\end{eqnarray}
Here the mass of $S_i$ is given by $m_{Si}$. The particle $S_i$ can obtain a mass
after the $B-L$ symmetry breaking as one can see from the interaction in
Eq.~(\ref{HP0}).
We use the Casas-Ibarra parametrization to express the Yukawa
matrix with the constraint of neutrino masses and
mixing~\cite{Casas:2001sr}. Then the product of
Yukawa matrix is written as 
\begin{equation}
y_{N}^{\dag}y_{\nu}=\sqrt{\Lambda}^{-1}C\sqrt{\hat{m}_{\nu}}U_{\mathrm{PMNS}}^{\dag},
\end{equation}
where the matrix $\Lambda$ is defined as
$(\Lambda)_{ij}=\Lambda_i\delta_{ij}$, $C$ is a complex orthogonal
matrix which satisfies $C^TC=1$, $\hat{m_\nu}$ is the diagonalized
active neutrino mass matrix and $U_{\mathrm{PMNS}}$ is the
Pontecorvo-Maki-Nakagawa-Sakata matrix. 
We need $\mathcal{O}(1)$ Yukawa couplings in order to produce the
proper DM relic density as will be discussed later. This corresponds to
$\Lambda\sim m_\nu\sim10^{-10}~\mathrm{GeV}$. 
In addition, for sum of active neutrino masses, the limit of $\sum
m_{\nu}<0.933~\mathrm{eV}$ at 95\% confidence level is imposed from the cosmological
observation~\cite{relic-2}.

Third, Lepton Flavor Violation (LFV) should be taken into account. 
The most stringent constraint comes from the LFV process $\mu\to
e\gamma$. Note that the LFV process $\mu\to3e$ would give a stronger constraint
when the mass difference between the right-handed neutrino and charged
scalar $\eta^+$ is sufficiently large~\cite{Toma:2013zsa}. The Branching Ratio (Br) of the process 
$\ell_{\alpha} \to \ell_{\beta}\gamma~(\alpha,\beta=e,\mu,\tau)$ 
is given by 
\begin{equation}
\mathrm{Br}\left(\ell_\alpha\to\ell_\beta\gamma\right)=
\frac{\alpha_{\mathrm{em}}\left|\left(y_{\nu}y_\nu^\dag\right)_{\alpha\beta}\right|^2}
{768\pi G_F^2m_\eta^4}
\mathrm{Br}\left(\ell_\alpha\to\ell_\beta\nu_\alpha\overline{\nu_\beta}\right),
\label{eq:lfv}
\end{equation}
where the right-handed neutrino masses are neglected. 
The latest limit for $\mu\to e\gamma$ is given by MEG
experiment~\cite{meg2} as
\be
\mathrm{Br}(\mu \to e \gamma)<5.7\times 10^{-13},
\label{constraints}
\ee
at 90\% confidence level.
For example, if the matrix $y_N$ is diagonal, the constraint of $\mu\to
e\gamma$ imposes that the orthogonal matrix $C$ should be almost unit
matrix $C\sim1$. In other words, we can see from the 
Casas-Ibarra parametrization that the product of the Yukawa matrix
$y_\nu y_\nu^\dag$ becomes almost diagonal since the PMNS matrix is
cancelled. Thus it does not contribute to 
any $\ell_\alpha\to \ell_\beta\gamma$ processes.

\section{Dark Matter Relics}
\label{sec3}
\begin{figure}[t]
\includegraphics[scale=0.7]{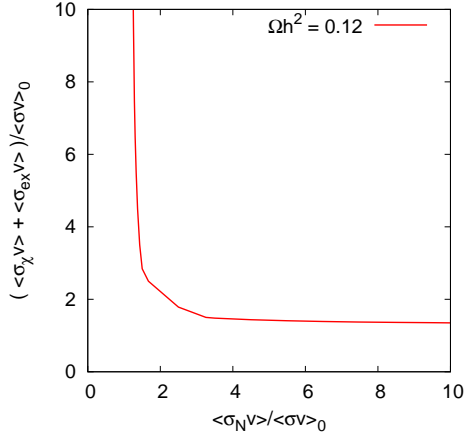}
\caption{$\langle\sigma{v}\rangle_0$ is the typical scale of
 annihilation cross section $2.0\times10^{-26}~[\mathrm{cm^3/s}]$.}
\label{fig:1}
\end{figure}

We have some DM candidates with odd under $\mathbb{Z}_2$ parity. They
are the right-handed neutrinos $N_i^c$, singlet fermion $S_i$ and neutral
component of $\eta$. It is natural to choose the lightest right-handed
neutrino as a DM since the mass is
generated at one-loop level and lighter than the other candidates. 
Hereafter we call the DM as $N_1$ with the mass $m_{N_1}$. 
In addition to the right-handed neutrino DM, we have an extra DM
candidate $\chi$. 
This is because after the breaking of the global $B-L$ symmetry, we still have remnant
discrete $\mathbb{Z}_6$ symmetry under which $\chi$ is odd. This
guarantees the stability of $\chi$. 
The lighter one of $\chi_R$
and $\chi_I$ can be the second DM, and we assume $\chi_R$ is DM. 
Thus we have two component DM of $N_1$ and $\chi_R$. 
The assumption of the mass hierarchy $m_{N_1}<m_{\chi_R}$ is reasonable from
the mass generation mechanism.
The DM $\chi_R$ can annihilate into the other DM (right-handed neutrino), but cannot
decay into the SM particles with the renormalizable interactions. 
They cannot be
taken care independently when one computes each relic density since
one DM annihilates into the other DM. 
The set of Boltzmann equations is written as 
\begin{eqnarray}
\frac{dn_N}{dt}+3Hn_N&=&
-\langle\sigma_{N}{v}\rangle\left(n_N^2-{n_N^{\mathrm{eq}}}^2\right)
+\langle\sigma_{\mathrm{ex}}{v}\rangle\left[n_\chi^2-\left(\frac{n_\chi^{\mathrm{eq}}}
{n_N^{\mathrm{eq}}}\right)^2n_N^2\right],\\
\frac{dn_\chi}{dt}+3Hn_\chi&=&
-\langle\sigma_{\chi}{v}\rangle\left(n_\chi^2-{n_\chi^{\mathrm{eq}}}^2\right)
-\langle\sigma_\mathrm{ex}{v}\rangle\left[n_\chi^2-\left(\frac{n_\chi^{\mathrm{eq}}}
{n_N^{\mathrm{eq}}}\right)^2n_N^2\right],
\end{eqnarray}
where the time of universe is expressed by $t$, $n_N$ and $n_\chi$ are the
number density of $N_1$ and $\chi_R$ respectively. 
The thermally averaged annihilation cross section into all channels is written as
$\langle\sigma_{N}{v}\rangle$ for $N_1$. 
For $\chi_R$, the total cross section into the SM particles is written  
by $\langle\sigma_{\chi}{v}\rangle$, and $\langle\sigma_{\mathrm{ex}}{v}\rangle$
implies the DM exchange process $\chi_R\chi_R\to NN$. 
If $\langle\sigma_{\mathrm{ex}}{v}\rangle$ is negligible compared with
$\langle\sigma_{\chi}{v}\rangle$, the simultaneous Boltzmann 
equation becomes independent of each other, and the total relic density
should be a sum of $N_1$ and $\chi_R$ : $\Omega_{N_1}h^2+\Omega_{\chi_R}h^2$. 
If $\langle\sigma_{\mathrm{ex}}{v}\rangle$ is a main channel of $\chi_R$
annihilation, the most of $\chi_R$ 
annihilates into the other DM $N_1$, but the relic density of $N_1$ almost does not
depend on the DM exchange process because $N_1$ is still in thermal
equilibrium when $\chi_R$ is frozen-out. 
Therefore the effect of the DM exchange process is small, and the DM
system can be treated as two independent DM as a good approximation. 
We have checked it numerically and the fact that effect of
semi-annihilation of DM is typically a few percent supports our
result~\cite{Belanger:2013oya}. 
The contours of satisfying $\Omega_{N_1} h^2+\Omega_{\chi_R} h^2=0.12$
are shown in Fig.~\ref{fig:1}. The x-axis and y-axis are the total cross section of $N_1$
and $\chi$ respectively. 

There are two channels for $N_1$ annihilation, which are 
$N_1N_1\to\ell\overline{\ell},\nu\nu$. 
The annihilation cross section for
$\ell\overline{\ell}$ and $\nu\nu$ are given as follows 
in the leading power of the DM relative velocity $v_\mathrm{rel}$:
\begin{eqnarray}
\sigma{v}_{\mathrm{rel}}\left(N_1N_1\to\ell\overline{\ell}\right)
&=&
\frac{\left[\left(y_{\nu}y_{\nu}^\dag\right)_{11}\right]^2}{48\pi
m_{N_1}^2}
\frac{m_{N_1}^4\left(m_{N_1}^4+m_\eta^4\right)}{(m_{N_1}^2+m_\eta^2)^4}
v_{\mathrm{rel}}^2,
\label{eq:ann-nc1}\\
\sigma{v}_{\mathrm{rel}}\left(N_1N_1\to\nu\nu\right)
&=&
\frac{\left[\left(y_{\nu}y_{\nu}^\dag\right)_{11}\right]^2}{24m_{N_1}^2}
\frac{m_{N_1}^4(m_{N_1}^4+m_0^4)}{(m_{N_1}^2+m_0^2)^4}v_{\mathrm{rel}}^2,
\label{eq:ann-nc2}
\end{eqnarray}
where $m_0^2$ is average mass between $m_{\eta_R}^2$ and
$m_{\eta_I}^2$ and they are assumed to be degenerate. 
Such a small mass difference is required to obtain a proper neutrino
mass scale as have discussed in ref.~\cite{Kajiyama:2013rla}. 
Note that for the annihilation into neutrinos the factor
2 larger than for the charged leptons because of Majorana property of neutrinos. 
The mass matrix of the right-handed neutrinos $m_N$ is generated at one
loop level and the expression is found as
\begin{equation}
\left(m_{N}\right)_{ij} = \sum_{k}\frac{(y_N)_{ik}(y_N)_{jk}m_{Sk}}{(4\pi)^2}
\left[\frac{m_{\chi_{R}}^2}{m_{\chi_{R}}^2-m_{{Sk}}^2}\ln\left(\frac{m_{\chi_{R}}^2}{m_{{Sk}}^2}\right)
-\frac{m_{\chi_{I}}^2}{m_{\chi_{I}}^2-m_{{Sk}}^2}\ln\left(\frac{m_{\chi_{I}}^2}{m_{{Sk}}^2}\right)\right].
\label{eq:numass}
\end{equation}
The DM and the mediator $\eta$ masses should be
$10\lesssim m_N\lesssim60~\mathrm{GeV}$ and
$100\lesssim m_\eta,m_0\lesssim300~\mathrm{GeV}$ to reproduce the  
correct relic density of the observed value $\Omega
h^2\approx0.12$~\cite{relic-2}. 
Otherwise the right-handed neutrino DM is
overproduced since the cross section becomes too small as one can see from
Eq.~(\ref{eq:ann-nc1}) and (\ref{eq:ann-nc2}). 
We also should note that there is the constraint from slepton 
search in SUSY models via the decay into lepton and missing energy at
LHC~\cite{ATLAS:2012zka, Aad:2012pxa, CMS:aro}. In our case, the charged
scalar $\eta^+$ has a similar properties with sleptons.  
The constraint on the mass of slepton is roughly $m_{\tilde{\ell}}\gtrsim270~\mathrm{GeV}$. 
Although this is not needed to be fully considered in our
case, one minds such the matter. 

For $\chi_R$ annihilation, there are five channels: $\chi_R\chi_R\to
hh,\: ZZ, W^+W^-,\: f\overline{f},\: GG$. Each cross section is
written by
\begin{eqnarray}
\sigma{v}(\chi_R\chi_R\to ZZ)
&=&
\frac{g_2^2m_Z^2}{4\pi s}\sqrt{1-\frac{4m_Z^2}{s}}
\left[3-\frac{s}{m_Z^2}+\frac{1}{4}\left(\frac{s}{m_Z^2}\right)^2\right]\nonumber\\
&&\times\left|\frac{\mu_{\chi\chi h}\cos\alpha}{s-m_h^2+im_h\Gamma_h}+
\frac{\mu_{\chi\chi H}\sin\alpha}{s-m_H^2+im_H\Gamma_H}\right|^2,
\end{eqnarray}
\begin{eqnarray}
\sigma{v}(\chi_R\chi_R\to WW)
&=&
\frac{g_2^2m_W^2}{2\pi s}\sqrt{1-\frac{4m_W^2}{s}}
\left[3-\frac{s}{m_W^2}+\frac{1}{4}\left(\frac{s}{m_W^2}\right)^2\right]\nonumber\\
&&\times\left|\frac{\mu_{\chi\chi h}\cos\alpha}{s-m_h^2+im_h\Gamma_h}+
\frac{\mu_{\chi\chi
H}\sin\alpha}{s-m_H^2+im_H\Gamma_H}\right|^2,
\end{eqnarray}
\begin{eqnarray}
\sigma{v}(\chi_R\chi_R\to f\overline{f})
&=&
\frac{y_f^2}{2\pi}\left(1-\frac{4m_f^2}{s}\right)^{3/2}
\left|\frac{\mu_{\chi\chi h}\cos\alpha}{s-m_h^2+im_h\Gamma_h}+
\frac{\mu_{\chi\chi H}\sin\alpha}{s-m_H^2+im_H\Gamma_H}\right|^2,
\end{eqnarray}
\begin{eqnarray}
\sigma{v}(\chi_R\chi_R\to hh)
&=&
\frac{1}{64\pi^2 s}\int 
\left|\frac{12\mu_{\chi\chi h}\mu_{hhh}}{s-m_h^2+im_h\Gamma_h}
+\frac{4\mu_{\chi\chi H}\mu_{hhH}}{s-m_H^2+im_H\Gamma_H}
\right.\nonumber\\
&&\qquad\qquad\left.
+\lambda_{10}\cos^2\alpha+\lambda_{12}\sin^2\alpha+\frac{4\mu_{\chi\chi
h}^2}{t-m_{\chi_R}^2}+\frac{4\mu_{\chi\chi h}^2}{u-m_{\chi_R}^2}
\right|^2d\Omega,
\end{eqnarray}
\begin{eqnarray}
\sigma{v}(\chi_R\chi_R\to GG)
&=&
\frac{1}{16\pi^2 s}\int 
\left|\frac{\mu_{\chi\chi h}\sin\alpha}{s-m_h^2+im_h\Gamma_h}
\frac{s}{v'}
-\frac{\mu_{\chi\chi H}\cos\alpha}{s-m_H^2+im_H\Gamma_H}\frac{s}{v'}
\right.\nonumber\\
&&\qquad\quad\qquad\left.
+\frac{\sqrt{2}m_5}{v'}-\frac{2m_5^2}{t-m_{\chi_I}^2}-\frac{2m_5^2}{u-m_{\chi_I}^2}
\right|^2d\Omega,
\end{eqnarray}
where $s$, $t$, $u$ are the Mandelstam variables, and the cubic
couplings $\mu_{\chi\chi h}$, $\mu_{\chi\chi H}$, $\mu_{hhh}$ and $\mu_{hhH}$ are
given by
\begin{eqnarray}
\mu_{\chi\chi
 h}&=&-\frac{m_5}{\sqrt{2}}\sin\alpha+\frac{\lambda_{10}}{2}v\cos\alpha-\frac{\lambda_{12}}{2}v'\sin\alpha, \label{eq:cch}\\
\mu_{\chi\chi
 H}&=&\frac{m_5}{\sqrt{2}}\cos\alpha+\frac{\lambda_{10}}{2}v\sin\alpha+\frac{\lambda_{12}}{2}v'\cos\alpha, \label{eq:ccH}\\
\mu_{hhh}&=&
\lambda_1v\cos^3\alpha-\lambda_6v'\sin^3\alpha
+\frac{\lambda_7}{2}v\sin^2\alpha\cos\alpha-\frac{\lambda_7}{2}v'\sin\alpha\cos^2\alpha,\\
\mu_{hhH}&=&
3\lambda_1v\sin\alpha\cos^2\alpha+3\lambda_6v'\sin^2\alpha\cos\alpha\nonumber\\
&&+\frac{\lambda_7}{2}v\sin^3\alpha-\lambda_7v\sin\alpha\cos^2\alpha
-\lambda_7v'\sin^2\alpha\cos\alpha+\frac{\lambda_7}{2}v'\cos^3\alpha.
\end{eqnarray}

As have discussed in ref.~\cite{Kajiyama:2013rla}, we need a large mass difference
between $\chi_R$ and $\chi_I$ in order to obtain a proper scale of active
neutrino masses. 
Hence the parameter relation is roughly estimated as 
\begin{equation}
\frac{m_5v'}{m_{\chi_R}^2}\gtrsim \mathcal{O}(1).
\label{m5-chi}
\end{equation}
The origin of the mass difference is the cubic coupling $m_5$. Thus we
can see that the cubic 
couplings $\mu_{\chi\chi h}$ and $\mu_{\chi\chi H}$ tend to be large
compared with the other couplings. In this case, the annihilation channels into gauge bosons ($ZZ$ and
$WW$) become dominant over the other channels because of the
longitudinal mode of the gauge bosons unless $\sin\alpha$ is extremely small.
The cross section is
roughly $\sigma{v}\sim10^{-24}~\mathrm{cm^3/s}$ when
$\sin\alpha\sim1$. 
As we will discuss later, such a large mixing angle is
excluded by direct detection of DM and the invisible decay
mode of the SM-like Higgs. 
In this case, the most of the DM 
$\chi_R$ disappears at the early universe and only the right-handed neutrino
DM remains. On the contrary, the annihilation channels into the Higgs and Goldstone
boson become dominant when the mixing is small such as $\sin\alpha\lesssim0.01$.
The DM exchange channel $\chi_R\chi_R\to N_1N_1$ also may be a leading channel. 
The cross section of the process $\chi_R\chi_R\to N_1N_1$ is found as
\begin{equation}
\sigma_{\mathrm{ex}}{v}_{\mathrm{rel}}\left(\chi_R\chi_R\to N_1N_1\right)
\!\approx\!
\sum_{i} \left[ \frac{(y_N)_{1i}^4}{8\pi m_{\chi_R}^2}\frac{\mu_i}{(1+\mu_i)^2}
-
\frac{(y_N)_{1i}^4}{24\pi m_{\chi_R}^2} 
\frac{\mu_i(1+3\mu_i)}{(1+\mu_i)^4}v_{\rm rel} ^2 \right]
\label{eq:tansit},
\end{equation}
where $\mu_i=m_{\chi_R}^2/m_{Si}^2$.
Notice here that the above cross section is the massless limit of
the final state particles.

Next we discuss detectability of the two DM candidates.
For the case of the scalar DM, it would be possible to detect it by
direct search if the cubic or quartic couplings in the scalar potential are
$\mathcal{O}(1)$\footnote{The right-handed neutrino DM also may be
detected through one loop photon exchange interaction
if the Yukawa matrix $y_\nu$ is complex and the mass is degenerate with
the second right-handed neutrino~\cite{Schmidt:2012yg}.}. The Higgs
exchange is a primary channel because the scalar 
DM does not have direct interactions with the SM 
particles except the Higgs potential. Thus $\chi_R$ is so-called Higgs
portal DM~\cite{Silveira:1985rk, Kim:2008pp, Hambye:2008bq,
Mambrini:2011ik, Raidal:2011xk, Okada:2010wd, Baek:2012se, Baek:2013dwa}. 
Since the term with $m_5$ is dominant in Eq.~(\ref{eq:cch}) and
Eq.~(\ref{eq:ccH}), the spin independent elastic scattering cross 
section with proton is written by 
\bea
\sigma_{p\mbox{-}\chi_R} \approx \frac{c}{8\pi}\frac{
m_p^4m_5^2\sin^22\alpha}{(m_{\chi_R}+m_p)^2 
v^2}\left(\frac{1}{m_h^2}-\frac{1}{m_H^2}\right)^2, 
\label{eq:sigma_p}
\eea
where $m_p=938$ MeV is the proton mass and $c\approx0.079$ is a
coefficient that is determined by the lattice
simulation~\cite{Corsetti:2000yq, Ohki:2008ff}~\footnote{When $m_h
\approx m_H$, there is cancellation between  
the two terms in~Eq. (\ref{eq:sigma_p}). And we can easily evade the direct 
detection bound when the mixing angle $\alpha \lesssim 0.4$ coming from the LHC Higgs searches~\cite{cancellation}.}. 
The stringent constraint can be obtained by LUX experiment that tells us
$\sigma_{p\mbox{-}\chi_R} \lesssim 7.6\times 10^{-46}$ cm$^{2}$ at $m_{\chi_R}\approx33$
GeV~\cite{xenonlux} where we implicitly assumed $\chi_R$ is dominant component of DM
in this estimation. 
When $m_p\ll m_{\chi_R}$ and $m_h\ll m_H$, the conservative limit is given by 
\begin{equation}
\frac{m_5\sin2\alpha}{m_{\chi_R}}\lesssim 0.11.
\end{equation}
It is not difficult to satisfy this relation because the mixing angle $\sin\alpha$ should be
sufficiently small in order to be dominant in two DM system. Otherwise
the DM $\chi_R$ becomes sub-dominant. 

When $m_h\gg m_H$, further smaller mixing angle $\sin\alpha$ is required
since the elastic cross section is enhanced by the light Higgs. 
However, it is interesting to consider such a light extra Higgs because
it is correlated with the additional
contribution of the Goldstone boson $G$ to the effective number of
neutrino species $N_{\mathrm{eff}}$~\cite{weinberg}. 
The discrepancy of the effective 
number of neutrino species $\Delta N_{\rm eff}$ has been reported by several
experiments such as Planck~\cite{relic-2}, WMAP9
polarization~\cite{Bennett:2012zja}, and ground-based
data~\cite{Das:2013zf, Reichardt:2011yv}, 
which tell us $\Delta N_{\rm eff}=0.36 \pm 0.34$ at the 68~\% confidence level. 
The Goldstone boson $G$ may contribute to the effective neutrino number $\Delta
N_{\mathrm{eff}}$ if the period of freezing out of the particle
is suitable. The appropriate era of freeze-out of the Goldstone boson is before muon
annihilation while the other SM particles are decoupled,
thus it corresponds to $T\approx m_\mu$ where 
$T$ is the temperature of the universe. 
The scattering of the Goldstone boson with the SM particles occurs through
the Higgs exchange. The interaction rate should be same order with the
Hubble parameter $H$ when $T\approx m_\mu$. 
From the rough evaluation of the reaction rate of $G$ and the Hubble
parameter, we obtain the condition
\bea
\frac{\sin^22\alpha(m_h^2-m_H^2)^2}{4(v v')^2} \frac{m_\mu^7 m_{\rm pl}}{m_h^4 m_H^4}
\approx 1,
\eea
where $m_{\rm pl} \approx 1.2 \times 10^{19}~{\rm GeV}$ is the Planck
mass and $m_\mu$ is muon mass. Typically the extra Higgs boson should be
light to satisfy this relation. 
As have discussed in ref.~\cite{weinberg}, the invisible decay mode
$h\to GG$ also constrains the mixing angle $\sin\alpha$. However, we
found that the constraint from the direct detection is stronger.  

\begin{figure}[t]
\begin{center}
\includegraphics[scale=0.7]{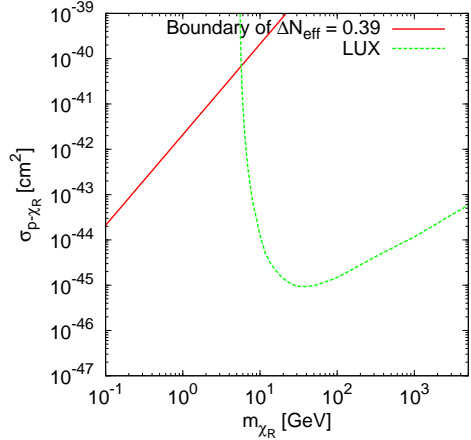}
\caption{The effective number of neutrinos and the constraint from LUX
 experiment.}
\label{fig:2}
\end{center}
\end{figure}

Combining with Eq.~(\ref{eq:sigma_p}) and (\ref{m5-chi}), the following
constraint on elastic cross section is obtained to get a certain value of $\Delta N_{\mathrm{eff}}$ 
\begin{equation}
\sigma_{p-\chi_R}\approx
\frac{c}{2\pi}\frac{m_p^4m_5^2v'^2}{m_{\chi_R}^2m_\mu^7m_{\mathrm{pl}}}
\gtrsim
\frac{c}{2\pi}\frac{m_p^4m_{\chi_R}^2}{m_\mu^7m_{\mathrm{pl}}}. 
\end{equation}
This requirement is shown with the limit of LUX experiment~\cite{xenonlux} in Fig.~\ref{fig:2}. 
The upper left region of the red line implies the region that $\Delta
N_\mathrm{eff}\approx0.39$ can be derived as ref.~\cite{weinberg}. Such a large elastic
cross section is obtained when the extra Higgs boson $H$ is quite
lighter than the SM-like Higgs $h$. 
The lower right region corresponds too fast deviation of the Goldstone
boson from thermal
bath and the contribution to $\Delta
N_{\mathrm{eff}}$ is negligible. 
As the figure, when we consider the case of $m_H\ll m_h$, we can get
upper bound on the DM mass 
$m_{\chi_R} \lesssim 5.5 \,{\rm GeV}$ from the LUX experiment~\cite{xenonlux}.
Therefore this result would contradict with the above discussion of
thermal relics of DM since we have assumed
$m_{N_1}<m_{\chi_R}$. However if $\chi_R$ is a sub-dominant component
DM, the constraint of LUX experiment is moderated. 
Or, we could also consider a light DM scenario such as $m_{\chi_R}<m_N$. 
Although we need a little fine-tuning for quartic couplings of the Higgs
potential is needed to obtain such a light Higgs mass, it is not
difficult to obtain sizable $\Delta N_{\rm
eff}\approx0.39$~\cite{weinberg,ibarra}.

\section{Differences from the original model}
\label{sec4}
This work includes some similar parts with our previous
paper~\cite{Kajiyama:2013rla}. 
Thus it is better that the main differences with our previous work
are made clear. 
In this paper, we suppose that the $U(1)_{B-L}$ symmetry is global,
unlike our previous paper~\cite{Kajiyama:2013rla} that has been
discussed in the local gauged symmetry. 
We do not need to introduce additional $\overline{S}$ as the
previous paper to cancel gauge anomaly, thus the new model is more
economical. 
More detail DM phenomenology with two-component DM was
discussed in this paper. 
In particular due to the global symmetry, we have a Goldstone boson $G$
that provides the feasibility of the observed discrepancy of the
effective number of neutrino species $\Delta N_{\rm 
eff}\approx0.39$ in a similar way of Weinberg
model~\cite{weinberg}. 

In addition to the above things, there is another expectation for the
global case. In ref.~\cite{Kajiyama:2013rla}, the DM candidates have
been $N_1$ and $\overline{S}$ which are fermion both. 
Then since their annihilation cross sections have been p-wave
suppressed, there has been no detectability for indirect detection. 
On the other hand, in this paper the scalar DM $\chi_R$ is included and
it has s-wave in general. 
The scalar DM $\chi_R$ has the mass of $\mathcal{O}(10)~\mathrm{GeV}$
from the view of the neutrino effective number $\Delta N_\mathrm{eff}$. 
Therefore the recently discussing gamma-ray excess below $10~\mathrm{GeV}$ would
be explained well by the scalar DM if the dominant annihilation channel
is $\chi_R\chi_R\to \tau\overline{\tau}$~\cite{Hooper:2013rwa, Boehm:2014hva}.

\section{Conclusions}
\label{sec5}
We have constructed a two-loop radiative  seesaw model with 
global $B-L$ symmetry at the TeV scale, which provides neutrino masses
with more natural parameters. 
Various phenomenological constraints such as S-T parameters, neutrino
masses and mixing and lepton flavor violation, stability of Higgs
potential have been taken into account. 
The Casas-Ibarra parametrization for the neutrino Yukawa matrix have been
used to describe the lepton flavor violating process $\mu\to
e\gamma$. 

We have studied the multi-component DM properties with fermion $N_1$
and scalar boson $\chi_R$. The mass of $N_1$ is generated at one-loop
level, thus $m_{N_1}<m_{\chi_R}$ is natural. 
The set of the Boltzmann equation for $N_1$ and $\chi_R$ is
solved simultaneously. 
For the relic density of $N_1$, a large Yukawa coupling $\mathcal{O}(1)$
is required to reduce the abundance appropriately. 
On the other hand, the relic density of $\chi_R$ depends on the Higgs
mixing angle $\alpha$. 
In case of large mixing angle $\alpha$, $\chi_R$
component DM can be sub-dominant since the cross section becomes quite large. 
In case of small $\alpha$, $\chi_R$ component can be
dominant. 

It would be also possible to detect the scalar DM by direct search through Higgs
exchange elastic scattering 
if the cubic or quartic
couplings in the scalar potential are sufficiently large, since an elastic
scattering occurs with quarks via Higgs exchange. 
The Higgs mixing angle $\alpha$ is extremely constrained
by the latest direct search experiment LUX, in particular when the extra
Higgs boson is much lighter than the SM-like Higgs. 

At the end, we have discussed the discrepancy of the effective number
 of neutrino species, $\Delta N_{\rm eff}$ between theory and experiments. We found that 
 light extra Higgs and small mixing angle is needed to obtain $\Delta
 N_{\rm eff} \approx 0.39$. Moreover, the scalar DM mass is quite
 limited as $m_{\chi_R} \lesssim 5.5$ GeV when we consider the current
 direct detection search of LUX. 


\section*{Acknowledgments}
This work is partly supported by NRF Research Grant  2012R1A2A1A01006053 (SB).
T.T. acknowledges support from the European ITN project (FP7-PEOPLE-2011-ITN,
PITN-GA-2011-289442-INVISIBLES). 

\end{document}